\documentclass{PoS}

\usepackage{stackrel}

\def\tr{{\rm tr}}
\def\Tr{{\rm Tr}}
\def\QCD{{\rm QCD}}
\def\R{{\mathbf R}}
\def\3{{\mathbf 3}}

\title{Thermodynamic characterizations of exotic and missing states~\thanks{We would like to thank Wojciech Broniowski for useful discussions and
collaboration on related topics. This work is supported by Spanish
Ministerio de Econom\'{\i}a y Competitividad and European FEDER funds
under contracts FIS2014-59386-P and FPA2015-64041-C2-1-P, Junta de
Andaluc\'{\i}a grant FQM-225, Basque Government grant IT979-16, and
Spanish Consolider Ingenio 2010 Programme CPAN (CSD2007-00042). The
research of E.M. is supported by the Universidad del Pa\'{\i}s Vasco
UPV/EHU, Bilbao, Spain, as a Visiting Professor.}
}

\ShortTitle{Thermodynamic characterizations of exotic and missing states}

\author{\speaker{Eugenio Meg\'{\i}as}$\;^{1}$
\\ Departamento de F\'{\i}sica Te\'orica, Universidad del
Pa\'{\i}s Vasco UPV/EHU, Apartado 644, E-48080 Bilbao, Spain\\ E-mail:
$^{1}$\email{eugenio.megias@ehu.eus} } \author{Enrique Ruiz
    Arriola$\;^{2}$, Lorenzo Luis Salcedo$\;^{3}$\\ Departamento de
    F\'{\i}sica At\'omica, Molecular y Nuclear and Instituto Carlos I
    de F\'{\i}sica Te\'orica y Computacional,  Universidad de Granada, E-18071 Granada, Spain \\ E-mail: $^{2}$\email{earriola@ugr.es},
    $^{3}$\email{salcedo@ugr.es} }

\abstract{Thermal shifts and fluctuations at finite temperature below
  the deconfinement crossover of QCD from hadronic matter to the
  Quark-Gluon Plasma provide a viable way to look for exotic and
  missing states with given quantum numbers in the hadronic
  spectrum. We study a realization of the Hadron Resonance Gas model
  in the light quark ($uds$) flavor sector of QCD to study: i) the
  entropy shifts, and ii) the fluctuations of electric charge, baryon
  number and strangeness; and extract from them the possible existence
  of missing and exotic states from a comparison with lattice
  data. The analysis of the entropy shift based on the free energy of
  the Polyakov loop suggests the existence of exotic hybrids $g q \bar
  q$ and $gqqq$.}

\FullConference{XVII International Conference on Hadron Spectroscopy and Structure - Hadron2017\\
		25-29 September, 2017\\
		University of Salamanca, Salamanca, Spain}

\begin{document}

\section{Introduction}
\label{sec:Introduction}

The partition function of Quantum Chromodynamics (QCD) can be written as
\begin{equation}
Z_{\QCD} = \Tr \, e^{-H_{\QCD}/T}= \sum_n e^{-E_n/T} \,, \qquad H_{\QCD} \psi_n = E_n \psi_n \,,
\end{equation}
where $\psi_n$ are all possible states in the spectrum. This is the
fundamental quantity for the study of its thermal properties,
illustrating the relation between the thermodynamics of the confined
phase of QCD and its hadronic spectrum below the phase transition,
where the states are color singlets. A particular realization of this
idea is given by the Hadron Resonance Gas (HRG) model, in which
hadrons (mesons and baryons) are considered as non-interacting and
point-like particles~\cite{Hagedorn:1984hz}. The existence of missing
states in QCD is related to the completeness of the hadronic spectrum,
which means not only missing hadrons but also possible exotic
states. Within the approach of the constituent quark models, there
have been some recent studies on the computation of the partition
function at low temperatures by considering a partonic
expansion around the vacuum~\cite{Megias:2013xaa}
\begin{equation}
Z_{\QCD} = Z_0 \cdot Z_{[q\bar{q}]} \cdot Z_{[qqq]} \cdot Z_{[\bar{q}\bar{q}\bar{q}]} \cdot Z_{[q\bar{q}g]} \cdot Z_{[q\bar{q}q\bar{q}]}  \cdots \,,
\end{equation}
which includes contributions of all kind of possible color singlet
states like hybrid states, tetraquarks, etc. In this contribution we
will review the ability of some thermal observables to help in the
characterization of the existence of missing and exotic states in the
spectrum of QCD.

\section{Entropy shifts and missing states}
\label{sec:Entropy}

Let us consider the physical situation in which one extra heavy charge belonging to representation $\R$ is added to the QCD vacuum. Then, the charge is screened by dynamical constituents to form color neutral states according to a specific pattern which depends on $\R$~\cite{Megias:2013xaa,Megias:2016onb}, and the energy of the states changes under the presence of the charge as~$E_n \to E_n^\R  =  \Delta_n^\R + m_\R$, where $\Delta_n^\R$ remains finite in the limit $m_\R \to \infty$. The Polyakov loop operator in representation~$\R$ is~$\tr_\R \, \Omega (\vec r) = \tr_\R \, e^{i A_0 (\vec r)/T}$. It has been argued in~\cite{Megias:2016onb} that the vacuum expectation value of the Polyakov loop is related to the so-called {\it free energy shift}, $\Delta F_\R$, due to the presence of a charge $\R$ in the medium, as
\begin{equation}
L_\R \equiv \langle \tr_\R \, \Omega (0) \rangle = \frac{Z_\R}{Z_{\mathbf 0}}= e^{-\Delta F_\R/T} = \frac{\frac{1}{2}\sum_n e^{-\Delta_n^\R/T}}{1+ \cdots} \,.
\end{equation}
The last equality of this formula was firstly introduced in~\cite{Megias:2012kb} for $\R = \3$, and constitutes a hadronic representation of the Polyakov loop in terms of color singlet states with one heavy quark. This means that this formula can be used to count states in the spectrum of QCD, like in the usual HRG model for the equation of state.
\begin{figure*}[htb]
\begin{tabular}{cc}
\includegraphics[width=70mm]{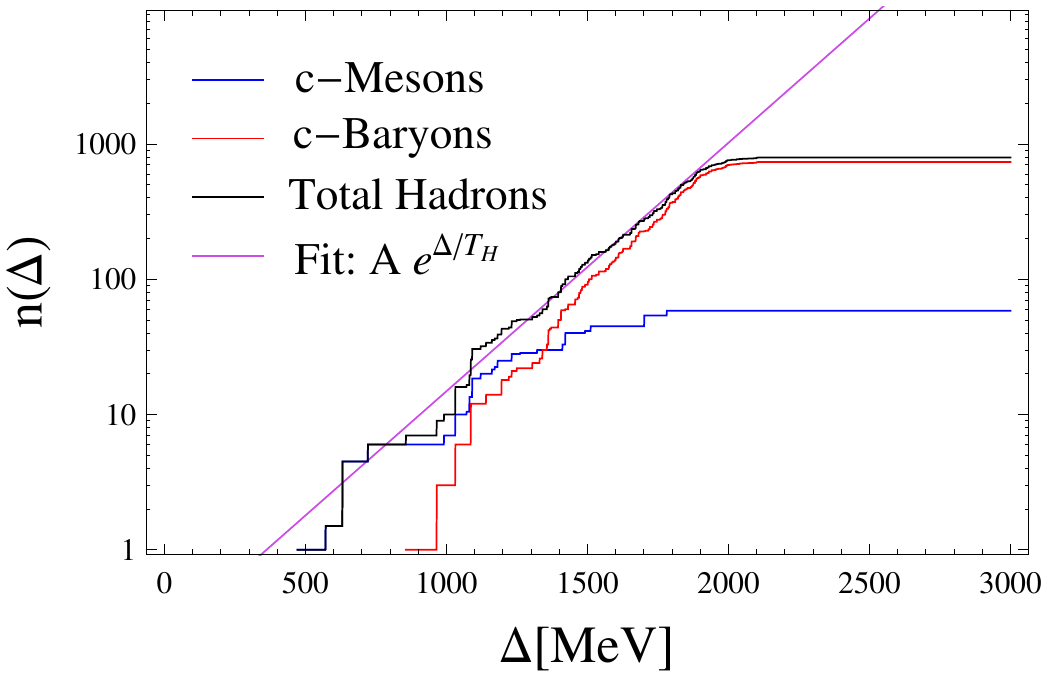} &
\includegraphics[width=70mm]{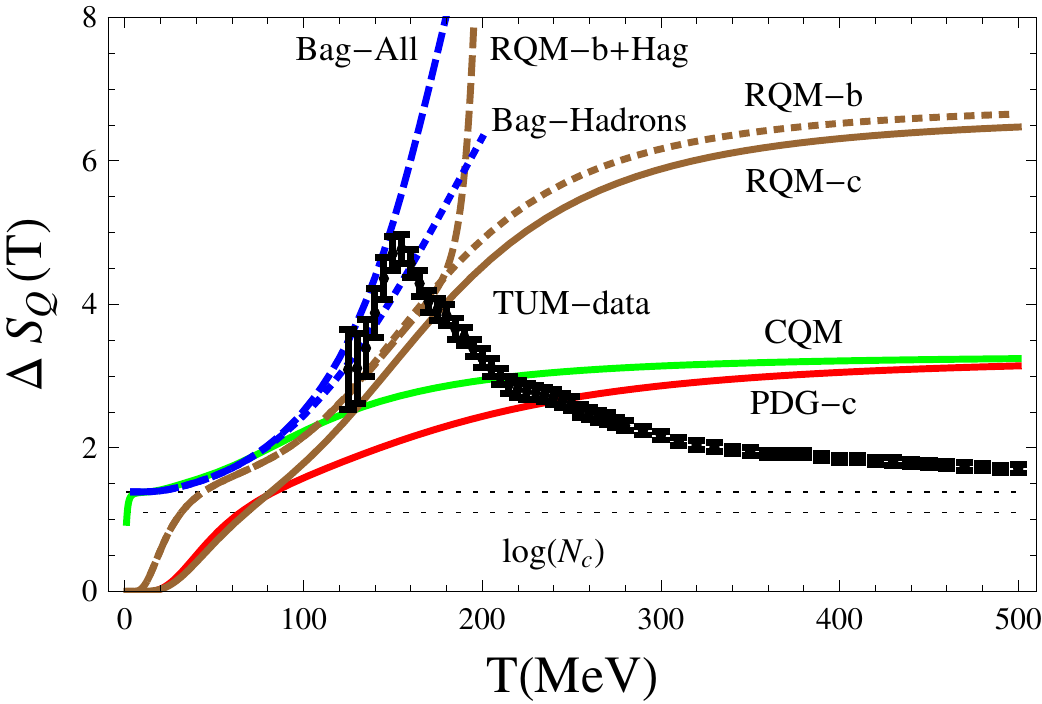} \\
\end{tabular}
\caption{Left panel: Cumulative number $n$ as a function of the shifted hadron
  mass~$\Delta$ (in MeV), for the hadron spectrum with a $c$ quark and light dynamical quarks $(uds)$ computed in the RQM~\cite{Godfrey:1985xj}. Right panel: Entropy shift as a function of
  temperature (in MeV). We display as dots the lattice data from Ref.~\cite{Bazavov:2016uvm}.}
\label{fig:entropy_shift}
\end{figure*}
The Polyakov loop is affected by a renormalization ambiguity $L_\R = e^{c/T} \cdot \tilde L_\R$, so that it is convenient to work with the entropy shift~\cite{Megias:2016onb}, defined as 
\begin{equation}
\Delta S_\R(T)= -\partial_T \Delta F_\R(T) \,.
\end{equation}
 It is displayed in Fig.~\ref{fig:entropy_shift} (left) the cumulative number of heavy-light states $n(\Delta) = \sum_n \Theta ( \Delta - \Delta_n) \sim e^{\Delta/T_{H,L}}$, and in Fig.~\ref{fig:entropy_shift} (right) the entropy shift as a function of the temperature. Notice that the states of the PDG are not enough to describe the lattice data of Ref.~\cite{Bazavov:2016uvm}. The result obtained from the spectrum of heavy-light mesons, $[Q\bar q]$, and baryons, $[Qqq]$, with the Relativistic Quark Model (RQM)~\cite{Godfrey:1985xj} are not enough either to describe the data, and only when including in the spectrum hybrid states, $[Q\bar qg]$ and $[Qqqg]$, computed with the MIT Bag model, one gets a satisfactory result.

\section{Fluctuations of conserved charges}
\label{sec:Fluctuations}

\begin{figure*}[htb]
\begin{tabular}{cc}
\includegraphics[width=60mm]{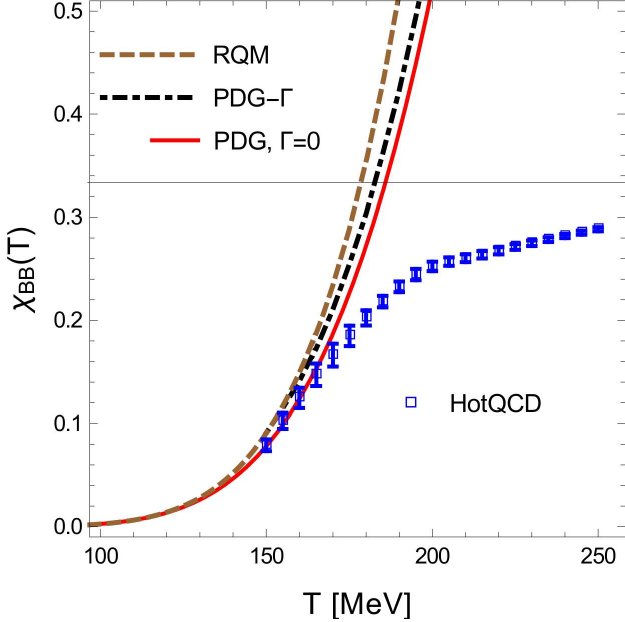} &
\hspace{1cm} \includegraphics[width=60mm]{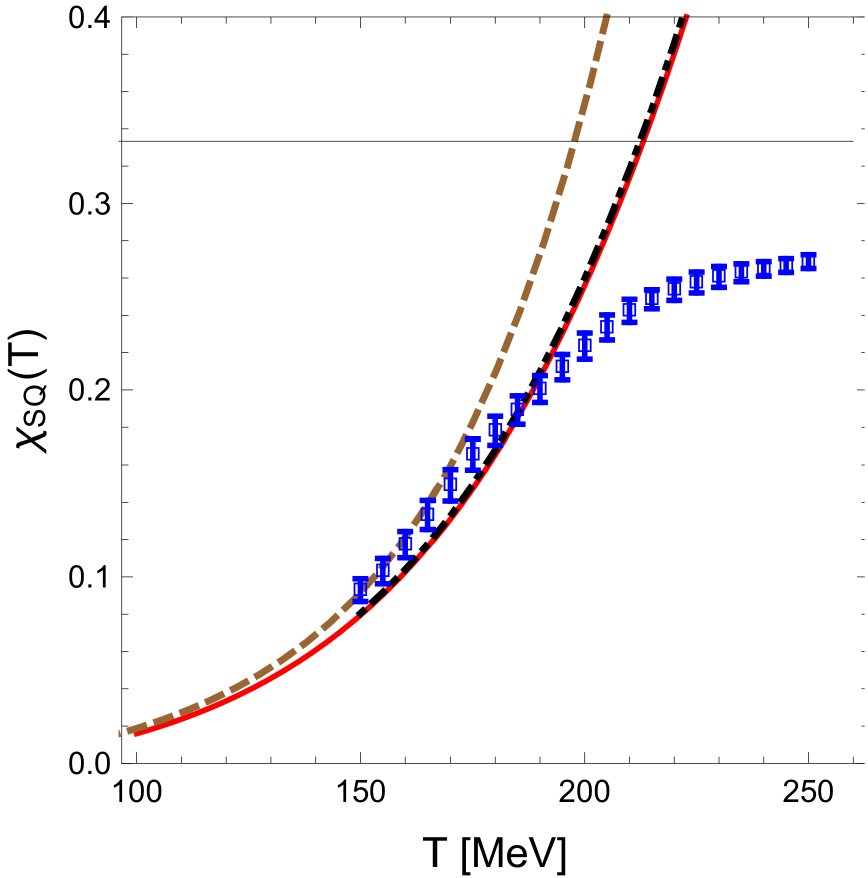} \\
\end{tabular}
\caption{Baryon, charge and strange susceptibilities from the HRG model with the PDG, PDG($\Gamma$) and RQM spectra. We display as dots the lattice data from Ref.~\cite{Bazavov:2012jq}.}
\label{fig:Fluctuations}
\end{figure*}

While the baryon number and other charges are conserved globally in QCD, their expectation values can be affected by local fluctuations due to thermal effects. These fluctuations have been related to the abundance of hadronic resonances in e.g.~Ref.~\cite{Jeon:1999gr}. Fluctuations of conserved charges, i.e. fulfilling $[Q_a,H]=0$, are a way of selecting given quantum numbers. They can be computed from the grand-canonical partition function, given by
\begin{equation}
Z = \textrm{Tr} \exp\bigg[ - \Big( H - \sum_a \mu_a Q_a \Big)/T \bigg] \,, \qquad \Omega = -T \log Z \,,
\end{equation}
where $\Omega$ is the corresponding potential. After differentiation one gets
\begin{equation}
-\frac{\partial \Omega}{\partial\mu_a} = \langle Q_a \rangle_T \,, \qquad  -T \frac{\partial^2 \Omega}{\partial\mu_a \partial\mu_b} = \langle \Delta Q_a \Delta Q_b \rangle_T \equiv VT^3 \chi_{ab}(T) \,,
\end{equation}
with $\Delta Q_a = Q_a - \langle Q_a \rangle_T$, and $\chi_{ab}$ are the susceptibilities. In the light quark $(uds)$ flavor sector the only conserved charges are the electric charge~$Q$, the baryon number~$B$, and the strangeness~$S$. Then, the susceptibilities in the HRG model are computed as
\begin{equation}
 \chi_{ab}(T) = \frac{1}{VT^3} \sum_{i,j \in {\rm Hadrons}} \!\! q_a^i q_b^j \langle \Delta N_i \Delta N_j \rangle_T \,, \qquad a,b \in \{ Q, B, S \} \,,
\end{equation}
where $Q_a = \sum_{i \in {\rm Hadrons}} q_a^i N_i$, and $q_a^i \in \{ Q_i, B_i , S_i\}$ are the charges of the $i$th-hadron corresponding to symmetry $a$. The averaged number of hadrons of type $i$ is
\begin{equation}
\langle N_i \rangle_T = V \int \frac{d^3k}{(2\pi)^3} \frac{g_i}{e^{E_{k,i}/T} - \xi_i} \,,  \qquad  E_{k,i} = \sqrt{k^2 + M_i^2} \,,
\end{equation}
with $\xi = \pm 1$ for bosons/fermions. We display in Fig.~\ref{fig:Fluctuations} our results for the susceptibilities $\chi_{BB}$ and $\chi_{SQ}$. While we find in general a good description of the lattice data~\cite{Bazavov:2012jq} for $T \lesssim 160$ MeV, one can see that the RQM seems to have too many baryonic states, but not too many charged states. This illustrates how fluctuations may help in the characterization of missing states in different sectors.

\section{Conclusions and outlook}
\label{sec:conclusions}
     
The thermal properties of QCD can be used to obtain information about missing states in the spectrum. The entropy shift due to a heavy quark placed in a thermal medium suggests that there are: i) conventional missing states~($[Q \bar q]$ and $[Qqq]$), and ii) hybrid states ($[Q\bar q g]$ and $[Qqqg]$). The fluctuations of conserved charges allow to study missing states in three different sectors: i) electric charge, ii) baryon number, and iii) strangeness. It would be interesting to extend this study with the correlation properties of conserved charges, and compare them with future lattice results.

\bibliographystyle{JHEP}
\bibliography{refs}

%

\end{document}